# Sensitivity of Ag/Al Interface Specific Resistances to Interfacial Intermixing.

A. Sharma,[1] N. Theodoropoulou,[1,2] Shuai Wang,[3] Ke Xia,[4] W.P. Pratt Jr.,[1] and J. Bass[1]

[1]Department of Physics, Michigan State University, East Lansing, MI 48824. USA
[2] Department of Physics, Texas State University, San Marcos, TX 78666.
[3]Institute of Physics, Chinese Academy of Sciences, P.O. Box 603, Beijing, 100190, China
[4] Department of Physics, Beijing Normal University, Beijing 100875, China

We have measured an Ag/Al interface specific resistance, $2AR_{Ag/Al}(111) = 1.4$ f$\Omega$m$^2$, that is twice that predicted for a perfect interface, 50% larger than for a 2 ML 50%-50% alloy, and even larger than our newly predicted 1.3 f$\Omega$m$^2$ for a 4 ML 50%-50% alloy. Such a large value of $2AR_{Ag/Al}(111)$ confirms a predicted sensitivity to interfacial disorder and suggests an interface ≥ 4 ML thick. From our calculations, a predicted anisotropy ratio, $2AR_{Ag/Al}(001)/2AR_{Ag/Al}(111)$, of more then 4 for a perfect interface, should be reduced to less than 2 for a 4 ML interface, making it harder to detect any such anisotropy.

In the past few years, our understanding of the interface specific resistance, $2AR_{M1/M2}$, for metals M1 and M2, with current flow perpendicular to the interface (Current-Perpendicular-to-Plane = CPP geometry), has greatly increased.[1] Here $2AR_{M1/M2}$ is twice the sample area A through which the CPP current flows, times the interface resistance $R_{M1/M2}$. For lattice matched metals with the same crystal structure and the same lattice parameter to within ~ 1%, measured values of $2AR_{M1/M2}$ agree surprisingly well with ones calculated with no adjustable parameters. Importantly for the present paper, in all five such cases studied so far, Co/Cu, Au/Ag, Fe/Cr, Pd/Pt, and Pd/Ir, the calculated values of $2AR_{M1/M2}$ for perfect (i.e., atomically flat and not intermixed) interfaces change only modestly (≤ 30%) when the interfaces consist of two monolayers (2ML) of a 50%-50% alloy.[1] Also, the experimental values of $2AR_{M1/M2}$ fall close to the predicted ones, lying mostly between those for perfect and 2ML alloyed interfaces, and in every case below the larger of the two values.[1]

Given these good agreements, our interest was piqued when Xu et al.[2] predicted for the lattice-matched pairs Ag/Al and Au/Al, two separate interesting results. While both Ag and Au form compounds with Al,[3] AuAl$_2$, called the 'purple plague',[4] is especially virulent, leading us to focus upon Ag/Al. First, for perfect interfaces, ref. [2] predicted a factor of four change with crystallographic orientation, $2AR_{Ag/Al}(111) = 0.64$ f$\Omega$m$^2$ vs $2AR_{Ag/Al}(001) = 2.82$ f$\Omega$m$^2$. Second, for 2ML of 50%-50% disorder, it predicted a 40% increase of $2AR_{Ag/Al}(111)$ to 0.92 f$\Omega$m$^2$, and a 15% decrease of $2AR_{Ag/Al}(001)$ to 2.37 f$\Omega$m$^2$.

These predictions stimulated us to try to measure $2AR_{Ag/Al}(111)$ to compare with predicted values for a wider range of intermixed ML, and also to extend the calculations to this same wider range. We expected interdiffusion between Ag and Al to lead to greater intermixing, because extrapolation of higher temperature diffusion data[5] to room temperature suggested that interdiffusion might occur at room temperature.[6] Our hope was three-fold. First, to try to measure $2AR_{Ag/Al}(111)$ in the presence of the expected intermixing. Second, to see if $2AR_{Ag/Al}(111)$ was as sensitive to intermixing as predicted. Third, to see if the intermixing was large enough to reduce the orientation dependence to where it would be harder to confirm.

We checked for evidence of intermixing using measurements of both AR and x-rays. For AR, we compare our values of $2AR_{Ag/Al}(111)$ with ones calculated for both perfect interfaces and disordered ones of several different thicknesses, and we also remeasured AR for selected samples after aging at room temperature for 2 years. We used x-rays both to check our initial growth orientation and layering, and then to look for evidence of changes in layering over time. Our value of $2AR_{Ag/Al}$ measured soon after sample deposition is much larger than the prediction for 2ML of a 50%-50% alloy. The values of AR for the selected samples also increased substantially over two years. We will take these two results as evidence for both early interfacial intermixing over a thickness of more than 2ML, and the predicted sensitivity of $2AR_{Ag/Al}$ to such intermixing. From x-rays, we concluded that intermixing was not strong enough to produce a uniform AgAl alloy, even after two years.

We estimated $2AR_{Ag/Al}$ using the technique of Ref. 7. This technique involves sputtering multilayers of Ag and Al having fixed total thickness of 360 nm, and dividing each sample into $n$ layers with equal thicknesses of Ag and Al—$t_{Ag} = t_{Al}$. To achieve the uniform CPP current flows needed to obtain reliable values of 2AR, these multilayers are sandwiched between ~ 1.1 mm wide, 150 nm thick, crossed superconducting Nb strips, as described in Ref. 8. The area A through which the CPP current flows is thus ~ 1.2 mm$^2$. To eliminate superconducting proximity effects on the multilayer, a 10 nm thick, ferromagnetic Co layer is placed between the multilayer and each Nb lead. No significant changes in total AR with magnetic fields from + 300 G to - 300 G show that these Co layers give negligible Giant Magnetoresistance.

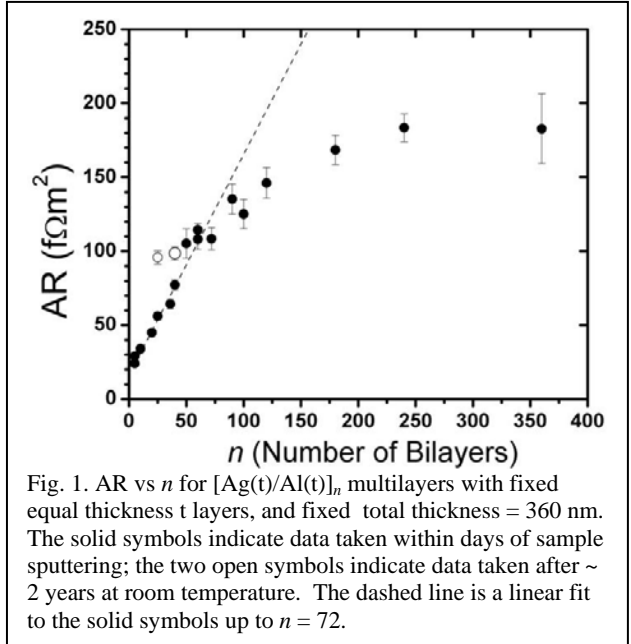

Fig. 1. AR vs $n$ for [Ag(t)/Al(t)]$_n$ multilayers with fixed equal thickness t layers, and fixed total thickness = 360 nm. The solid symbols indicate data taken within days of sample sputtering; the two open symbols indicate data taken after ~ 2 years at room temperature. The dashed line is a linear fit to the solid symbols up to $n$ = 72.

For Ag and Al layer thicknesses larger than the thickness of the intermixed interface, if we absorb all contributions from AgAl interfaces into 2AR$_{Ag/Al}$, AR should be given simply by:[6]

AR = 2AR$_{Nb/Co}$ + 2ρ$_{Co}$(10) + AR$_{Co/Al}$ + AR$_{Co/Ag}$ + ρ$_{Ag}$(180) + ρ$_{Al}$(180) - AR$_{Ag/Al}$+ $n$(2AR$_{Ag/Al}$),.   (1)

Here all layer thicknesses are in nm, and - AR$_{Ag/Al}$ occurs because each sample has only $2n - 1$ interfaces.

So long as Eq. 1 applies, plotting AR vs $n$ should give a straight line with slope 2AR$_{Ag/Al}$ and ordinate intercept equal to the sum of the first 7 terms. When the layer thicknesses approach that of the intermixed interface, AR should begin to saturate toward a value for an eventual 360 nm thick 50%-50% alloy plus the first four terms of Eq. (1). Fig. 1 shows such a plot for multilayers of the form Nb(150)/Cu(10)/Co(10)/[Ag(t)/Al(t)]$_n$/Co(10)/Cu(10)/Nb(150), with equal thicknesses t of Ag and Al and total [Ag/Al]$_n$ multilayer thickness of 360 nm (The Cu layers are included for clean growth—prior studies show that they become superconducting by the proximity effect with the Nb and don't affect 2AR$_{Nb/Co}$).[9] The solid symbols in Fig. 1, representing measurements of AR taken within days of sample preparation, are consistent with the expected form. Associating all of the linear growth with the interfaces, gives 2AR$_{Ag/Al}$(111) = 1.4 ± 0.2 fΩm$^2$ and intercept 20 ± 5 fΩm$^2$. This value of 2AR$_{Ag/Al}$ is about twice that predicted for a perfect interface, and 50% larger than predicted for 2 ML of a 50%-50% alloy.[2] As shown in Table I, however, it is close to the value that we predict for 4 ML of a 50%-50% alloy. A 4 ML interface thickness also agrees with the approximate saturation of AR at $n \approx 200$, translating to Ag and Al thicknesses of (360 nm)/400 = 0.9 nm = 4 ML.

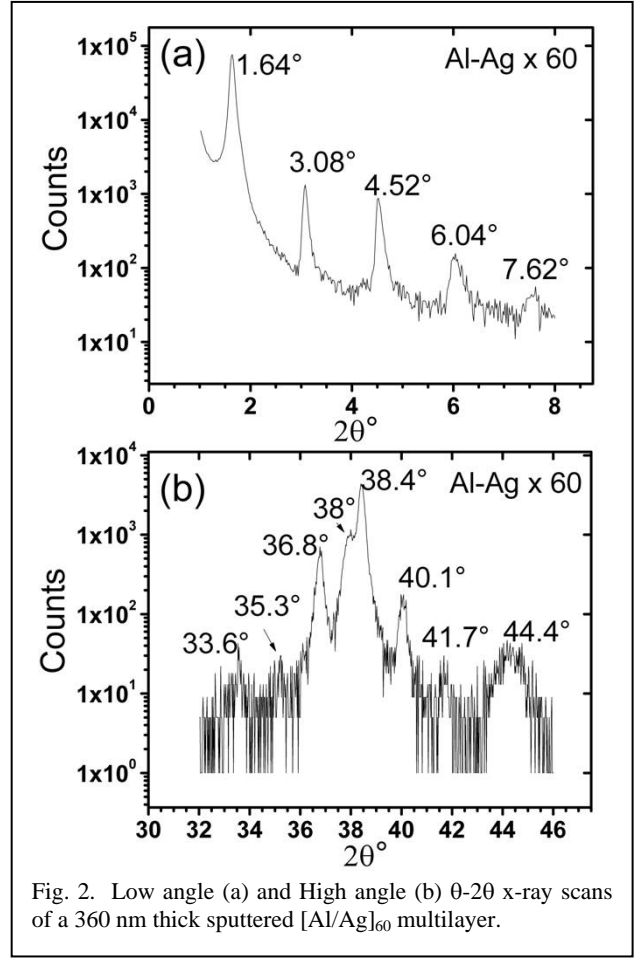

Fig. 2. Low angle (a) and High angle (b) θ-2θ x-ray scans of a 360 nm thick sputtered [Al/Ag]$_{60}$ multilayer.

From independent measurements of: 2AR$_{Nb/Co}$ = 6 ± 1 fΩm$^2$,[9] ρ$_{Co}$ = 62 ± 9 nΩm, ρ$_{Al}$ = 5 ± 1 nΩm, ρ$_{Ag}$ = 16 ± 4 nΩm, AR$_{Co/Ag}$ = 0.18 ± 0.02 fΩm$^2$,[10] and AR$_{Co/Al}$ = 5.5 ± 1 fΩm$^2$,[11] we estimate an intercept of 16 ± 3 fΩm$^2$. This value overlaps the experimental 20 ± 5 fΩm$^2$.

As noted above, extrapolations from higher temperature diffusion data[5] suggest that Ag can interdiffuse into Al at room temperature. We estimate bulk interdiffusion distances in a week of ~ 0.3 nm, and in two years of ~ 3 nm.[6] However, both values are uncertain by at least a factor of two. To test for possible long time interdiffusion, we remeasured two samples ($n$ = 25 and 40) in the linear region of distinct interfaces, after they aged at room temperature for about 2 years. The open symbols in Fig. 1 show increases in AR that give strong evidence of continuing interdiffusion between Ag and Al. However, neither value reaches the high $n$ 'saturation' limit that likely represents a nearly uniform AgAl alloy.

We used x-rays first to check the orientation of sputtered 200 nm thick Ag and Al films. We found the expected (111) peaks at d(Ag) = 0.236 ± 0.006 nm and d(Al) = 0.234 ± 0.006 nm.[5] We also looked for direct effects of interdiffusion on selected samples, both within a few days and after 2 year's aging. As examples, Fig. 2a shows a low angle θ-2θ x-ray scan, and Fig. 2b a high angle scan, for a Ag/Al $n = 60$ multilayer soon after preparation. The high angle central peaks in Fig. 2(b) agree with those for the 200 nm thick films. Both the low and high angle scans also show the 'satellite' peaks expected for layering. The low angle scan gives $d_{bilayer} = 5.9 \pm 0.3$ nm, in good agreement with the intended 6 nm. The 'initial' samples are clearly layered, but we cannot rule out some interdiffusion at the interfaces.

Given the changes in values of AR shown in Fig. 1 upon holding two samples ($n = 25$ and 40) at room temperature for ~ 2 years, we x-rayed the $n = 40$ sample after its new AR measurement. Multiple satellite peaks at both low and high angles showed that the sample was still layered. This result is consistent with the comment above that the changes in AR from filled to open symbols in Fig. 1 are due to more modest structural changes than complete intermixing of the Ag and Al.

We now turn to our calculations. Both the original calculations for Ag/Al,[2] and our extensions in Table I, assume that scattering in the bulk Al and Ag is fully diffuse, and that there is no coherence between scattering from adjacent interfaces. As shown in ref. [12], these two assumptions lead to a simple series resistor equation like Eq. 1 for either ballistic or diffuse scattering at the interface. The calculations assume a single lattice for both Al and Ag, with the lattice parameter, a, of Al to conform with ref. [2].. Following earlier work,[1,2] the electronic structures of Al and Ag are calculated using the local density approximation with Muffin Tin Orbitals (MTO) containing s, p, and d orbitals. The uncertainties in the calculated values of 2AR allow for uncertainty in the Fermi energy of 0.05 eV.[13] Table I compares our experimental value of 2AR(Ag/Al) with values calculated for a perfect interface and 50%-50% random alloys with thicknesses of 2, 4, and 6 ML (corresponding to ~ 0.5, 0.9, and 1.4 nm). Our measured value falls closest to the calculation for 4 ML. Intriguingly, while $2AR_{Ag/Al}(111)$ increases roughly linearly with interface intermixing thickness through 6 ML, $2AR_{Ag/Al}(001)$ has a minimum near 2 ML. This minimum is due to an initial opening by impurities of a partially blocked channel, after which more impurity scattering increases $2AR_{Ag/Al}(001)$

To summarize, measurements, made within days of sputtering, of AR vs $n$ for $[Ag(t)/Al(t)]_n$ multilayers with fixed total $t_T = n(2t)$, yield an Ag/Al interface resistance of $2AR_{Ag/Al}(111) = 1.4 \pm 0.2$ fΩm$^2$. This value is much larger than the no-free-parameter calculations for a perfect, unmixed interface, or for 2ML of a 50-50 alloy. It falls closest to the value $1.31 \pm 0.02$ fΩm$^2$ for 4 ML of a 50-50 alloy. From this agreement, we conclude that Ag/Al interfaces are significantly intermixed and that, as predicted,[2] $AR_{Ag/Al}(111)$ is sensitive to such intermixing. Further evidence for a large effect of intermixing comes from large increases in AR for two samples in the linear region of Fig. 2 upon holding at room temperature for ~ 2 years, coupled with the presence of both low and high angle x-ray satellites showing persistence of multilayer layering, instead of formation of a random alloy throughout the sample. From Table I, we see that our measured value of $2AR_{Ag/Al}(111)$ indicates a ≥ 4 ML thick intermixed interface. In such a case, the calculated values in Table I reduce the ratio $r = AR_{Ag/Al}(001)/AR_{Ag/Al}(111)$ from r > 4 for a perfect interface to r ≤ 2 if one can produce an (001) oriented sample.

Acknowledgments. This research was supported in part by US NSF grant DMR-08-04126. Ke Xia thanks NSF of China and MOST (No. 2006CB933000) of China.

References

[1] R. Acharyya, H.Y.T. Nguyen, R. Loloee, W.P. Pratt Jr., J. Bass, Shuai Wang, Ke Xia, Appl. Phys. Lett. **94**, 022112 (2009) and references therein.

[2] P.K. Xu, K. Xia, M. Zwierzycki, M. Talanana, P.J. Kelly, Phys. Rev. Lett. **96**, 176602 (2006).

[3] M. Hanson, Constitution of Binary Alloys', 2$^{nd}$ Ed., McGraw-Hill, New York, 1958.

[4] G.G. Harman, "Wire Bonding in Microelectronics", 2$^{nd}$ Ed., McGraw-Hill, 1997.

[5] H. Mehrer, N. Stolica, N.A. Stolwijk, 'Diffusion in Solid Metals and Alloys", ed. H. Meher, Landolt-Bornstein New Series, Group III, V 26, Springer, Berlin, 1990.

[6] A. Sharma Ph.D. thesis, Mich. State U. (2008) Unpublished.

[7] L.L. Henry, Q. Yang, W.-C. Chiang, P. Holody, R. Loloee, W.P. Pratt Jr., J. Bass, Phys. Rev. **B54**, 12336 (1996).

[8] S.F. Lee, Q. Yang, P. holody, R. Loloee, J.H. Hetherington, S. Mahmood, B. Ikegami, K. Vigen, L.L. Henry, P.A. Schroeder, W.P. Pratt Jr., J. Bass, Phys. Rev. **B52**, 15426 (1995).

[9] C. Fierz, S.-F. Lee, J. Bass, W.P. Pratt Jr., P.A. Schroeder, J. Phys. Cond. Matt. **2**, 9701 (1990).

[10] J.Bass and W.P. Pratt Jr., J. Magn. Magn. Mat. **200**, 274 (1999).

[11] A. Sharma, N. Theodoropoulou, T. Haillard, R. Acharyya, R. Loloee, W.P. Pratt Jr., J. Bass, Phys. Rev. **B77**, 224438 (2008).

[12] K.M. Schep, J.B.A.N. van Hoof, P.J. Kelly, G.E.W. Bauer, J.E. Inglesfield, Phys. Rev. **B56**, 10805 (1997).

[13] O.K. Andersen, Phys. Rev. **B2**, 883 (1970).

Table I, Experimental (exp) and calculated values of 2AR(Ag/Al) in units of fΩm$^2$. The only experimental value is for (111) oriented interfaces. Calculated values, for both (111) and (001) are for perfect interfaces (perf) and for 50%50% random alloys of thicknesses 2 ML, 4 ML, and 6 ML.

| Metals | Δa/a(%) | 2AR(exp) | 2AR(perf.) | 2AR(2ML) | 2AR(4ML) | 2AR(6ML) |
|---|---|---|---|---|---|---|
| Ag/Al(111) | 0.9 | 1.4± 0.2 | 0.64 ± 0.01 | 0.92 ± 0.01 | 1.31 ± 0.02 | 1.65 ± 0.03 |
| Ag/Al(001) | 0.9 |  | 2.82 ± 0.03 | 2.39 ± 0.04 | 2.50 ± 0.07 | 2.63 ± 0.04 |